# From Modal to Multimodal Ambiguities: a Classification Approach

[1]Maria Chiara Caschera, [2]Fernando Ferri, [3]Patrizia Grifoni

*[1,First Author] National Research Council - Institute of Research on Population and Social Policies (CNR-IRPPS), mc.caschera@irpps.cnr.it*

*[2,Corresponding Author] National Research Council - Institute of Research on Population and Social Policies (CNR-IRPPS), fernando.ferri@irpps.cnr.it*

*[3] National Research Council - Institute of Research on Population and Social Policies (CNR-IRPPS), patrizia.grifoni@irpps.cnr.it*

## Abstract

*This paper deals with classifying ambiguities for Multimodal Languages. It evolves the classifications and the methods of the literature on ambiguities for Natural Language and Visual Language, empirically defining an original classification of ambiguities for multimodal interaction using a linguistic perspective. This classification distinguishes between Semantic and Syntactic multimodal ambiguities and their subclasses, which are intercepted using a rule-based method implemented in a software module. The experimental results have achieved an accuracy of the obtained classification compared to the expected one, which are defined by the human judgment, of 94.6% for the semantic ambiguities classes, and 92.1% for the syntactic ambiguities classes.*

**Keywords:** *Human Computer Interaction, Language Ambiguity, Multimodal Language*

## 1. Introduction

In the last decade some critical aspects of the users' cognitive effort in Human Computer Interaction (HCI) processes, have been gradually shifted from users to computerized systems improving interaction naturalness involving the five human senses. Toward this direction, multimodal interaction enables combining different kinds of information, such as visual information involving images, texts, sketches and so on - with voice, gestures and other typical modalities of human-human interaction [1].

Oviatt [2] underlines the greater naturalness, of multimodal interaction systems respect to the conventional windows-icons menus-pointers (WIMP) interfaces. As examples of works to address naturalness issue, König et al. [3] provide an approach to support interaction design, while in [4] a generic framework that overcomes many difficulties associated with real world user behaviors analysis, e.g. dynamic real time analysis of multimodal information, is provided. The importance of human factors' role in multimodal interface design is analysed in [5].

Naturalness in multimodality can simplify the user's activity, but systems have to recognize and interpret complex inputs. Consequently, a natural and simplified interaction implies the improvement of systems complexity to manage typical problems of the human-human communication, such as ambiguity. An ambiguity arises when humans communicate visually and/or using the natural language and more than one interpretation is possible. Concerning ambiguity, studies on creative and communication activities have underlined that ambiguity can be regarded as both a positive and an active element in group communication processes. Ambiguity is frequently used in an active manner: to negotiate an intended meaning; to identify problems and their solutions; and, afterwards, to refine them [6], [7], [8], making processes more flexible and natural. Therefore, it is usually preferred to manage ambiguities by detecting the main features dealing with the disambiguation process, instead of preventing them. In the literature, works on ambiguities frequently focus on the problem of single modal recognition, e.g. in [9]. Our perspective focuses on ambiguities produced both by the propagation at a multimodal level of ambiguities arising by the recognition process of single modality, and by the combination of unambiguous modal information generating contrasting concepts at a multimodal level [10]. Therefore, considering the taxonomy on data fusion methods provided by Nigay and Coutaz [11] on multimodal systems, this paper focuses on ambiguities (at a multimodal level) that arise in systems based on syntactic and semantic fusion. This type of fusion requires a uniform representation of the inputs originated from different modalities. In order to provide a uniform representation, we adopt a linguistic approach to represent multimodal features and, then to classify multimodal ambiguities according to a linguistic point of view by a syntactic and semantic framework.

Figure 1 explains the Multimodal interaction process in its complexity and the role of the "Multimodal Ambiguity Classification", which is the core of this paper, in the more general interpretation process of multimodal inputs. As shown in Figure 1, the user multimodal inputs are processed by unimodal recognition modules (speech, handwriting, sketch, and so on), and the recognized signals for the various modalities are integrated (Fusion module) [12] and interpreted (Multimodal Interpreter module) according to a multimodal grammar [13]. The interpretation (for a single modality or a combination of different modalities) could not be unique, producing an ambiguity. If the Multimodal Interpreter produces more than one interpretation of the user's input, then the Multimodal Ambiguities Classification manages the ambiguous interpretation providing its classification. Then, information about the classified ambiguous input and the set of candidate interpretations are transferred to the Multimodal Ambiguities Solution module; to solve ambiguities it was adopted the module InteSe [14], which selects only one interpretation. The interpreted input will produce a response by the system (System Response Generation module). This response will be arranged in the fission step (Fission module) to be presented to the users.

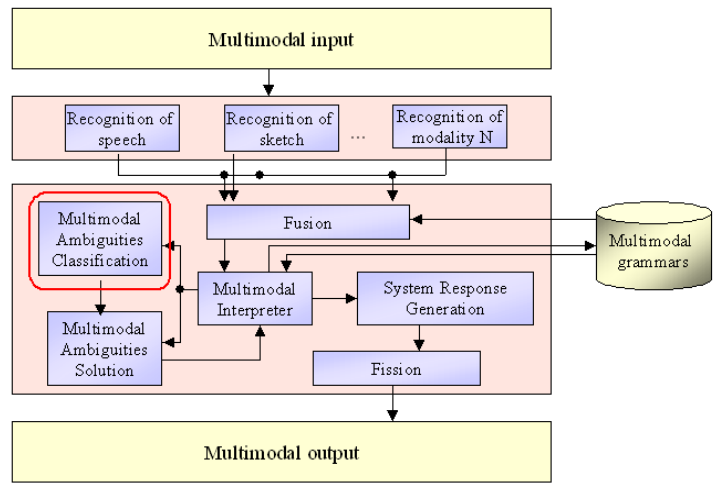

**Figure 1.** Steps of the multimodal interaction management

This paper discusses the classification step proposing a new classification that extends and reformulates the ambiguity classifications presented for Natural Language (NL) [15] and Visual Languages (VLs) [16] and evolves previous work on multimodal ambiguities [17].

The proposed method has been empirically defined (see section 4) using a set of 520 multimodal inputs (i.e. the multimodal sentences), 480 ambiguous and 40 un-ambiguous. Each class of ambiguity is identified by features described by a set of rules. The classification rules identify the features of the multimodal inputs and how they match the features of the different classes of ambiguities. The identification of a specific class of ambiguities enables to detect the features to be managed for simplifying the interpretation process and optimizing the resolution process. In particular, ambiguities can be connected to the structure (i.e. the syntax of a language) or/and to the semantic meaning of the input. If an ambiguity connected to the syntactic structure is detected, it is necessary to solve the syntactic problem. Similarly, when an ambiguity connected to the meaning of the multimodal input is detected, it is necessary to deal the semantics of the input. The resolution task is not discussed here because it is outside of the focus of this work.

The paper is organized as follows: section 2 presents an overview of the literature works on the classification of ambiguities in NL, VL; section 3 gives the notions of *Multimodal Attribute Grammar*, *Multimodal Sentence* and *Multimodal Language* used in the remaining sections; section 4 provides the multimodal ambiguities classification method; section 5 describes the Multimodal Ambiguity Classifier Module; section 6 shows the results of the evaluation tests of the Multimodal Ambiguity Classifier Module. Finally section 7 concludes the paper and presents future works.

## 2. Related work on Languages Ambiguities Classification

A communication process implies the sharing of common meanings for communication acts (such as a sentence for NL, or more generally a message) involving different actors. When there is a gap between the intended meaning of the sender's message and the meaning that the receiver gives to the message, then an error or an ambiguity could arise. A discussion on some among the most relevant studies on ambiguity dealing with NL, visual communication and VL, is presented in the next section.

### 2.1. Ambiguity classes in Natural and Visual Languages

Using NL involves the knowledge of words with their meanings (lexical, semantic), of the syntactic structure of sentences, and it implies to know how the different components are combined in order to assume larger meanings (compositional semantic). Therefore, un-ambiguous understanding of a sentence implies: 1) sharing a dictionary of words and their meanings; 2) a well known and common syntactic structure that enables to identify both, the syntactic role (i.e. a noun phrase, a verb, an adjective, an adverb, a pronoun, a preposition, etc.) assigned to each word in a sentence by the syntax of the grammar, and how the different words are grouped and connected with, and finally; 3) knowledge about the opportunely combined meanings of words. Berry et al. [15] introduced four classes of linguistic ambiguities for NL: lexical, syntactic, semantic and pragmatic ambiguities. These four classes are not mutually exclusive, because an ambiguous sentence may contain a combination of several of these classes.

A Lexical ambiguity appears when one word has more than one generally accepted meaning, for example in the case of homonymy or polysemy. Homonymy occurs when two words have the same representation but different meanings, while polysemy occurs when a word has different meanings connected to each other, and the process of extension of similar meanings clearly arises [18].

Syntactic ambiguity consists of an ambiguity arising from the way the sentence is structured. The syntactic structure of the sentence in NL can be represented by a parse tree [19] according to a given formal grammar. A parse tree, or concrete syntax tree, is an ordered and rooted tree obtained by analyzing a sequence of inputs in NL. The internal nodes are labeled by non-terminals of the grammar, and leaves by terminal symbols. A sentence in NL presents a syntactic ambiguity if and only if there is more than one parse tree for it. These ambiguities in Berry et al., [15] are classified in: analytic, attachment, coordination and gap ambiguities. Analytic ambiguities arise when some syntactic role in a sentence is not clearly identified, as for example in the sentence "Tibetan history teacher" where two possible interpretations are "(Tibetan history) teacher" and "history (Tibetan teacher)" [15]. An attachment ambiguity occurs when a set of words in a sentence can be legally attached to two different parts of the sentence; this is due to different syntactic structures of the sentence, similarly to the analytic ambiguity. In this case, a prepositional phrase or a relative clause can be legally attached to different parts of the sentence. An example of attachment ambiguity is given by the sentence "the police shot the rioters with guns" that can have two possible interpretations: "the police shot (the rioters) with guns" and "the police shot (the rioters with guns)" [15]. A coordination ambiguity arises when more than one conjunction, "and" or "or", is used in a sentence or when there is coordination with an adjective (e.g. "I saw John and Alex and Clare saw me"- this sentence can be read either as "I saw (John and Alex) and Clare saw me" or as "I saw John and (Alex and Clare) saw me"-). A gap ambiguity can arise if an element of the sentence - i.e. the elementary component of a sentence associated with a concept, expressed using NL in a given time interval and playing a role in the sentence - is omitted. It is defined by an ellipsis in a sentence, i.e. the omission of lexically or syntactically necessary components. As an example, the sentence "Perot knows a richer man than Trump" could have two different meanings: "Perot knows a man who is richer than Trump is" and "Perot knows a man who is richer than any man Trump knows" [15].

Semantic ambiguities appear when a sentence has more than one interpretation, even if neither lexical nor syntactic ambiguities appear in the sentence [15]. They could be produced by a coordination ambiguity, a scope ambiguity and a referential ambiguity. Coordination ambiguities have been already presented. A Scope ambiguity appears when there are quantifiers that can define different relations in the sentence, e. g. "All linguists prefer a theory" [15]; this sentence can be read either as "all linguists love the same one theory" or "each linguist loves a, perhaps different, theory". A Referential ambiguity is also a class of pragmatic ambiguities; it refers to the meaning of the word connected with the context, and in particular when there is an anaphora and it can be referred to two different words.

A pragmatic ambiguity appears when a sentence can have more than one meaning in a single context; this class is divided into: referential ambiguity, and deictic ambiguity, i.e. ambiguity that appears when pronouns and other grammatical elements have more than one reference in the discourse context. Referential ambiguity is on the boundary between semantic and pragmatic ambiguity. In fact, it can occur within a sentence or between a sentence and its discourse context.

A classification for VL, which is similar to the classification presented for ambiguities in NL, is provided by Futrelle [16]. The concept of ambiguity for visual information was widely implied in the cultural debate on the Gestalt theory [20]. Gestalt, with its idea that a figure and its background can exchange their roles according to the adopted point of view, has deeply influenced the cultural scenario during the 20th century. The literature on VLs provides some different definitions of ambiguity. Futrelle [16] distinguishes between lexical and syntactic ambiguities for VLs, with the meaning previously introduced for NL. Favetta and Aufaure-Portier [21] defined taxonomy of ambiguities for Visual GIS query languages. This taxonomy was developed and described using the formalism for VLs introduced by Bottoni et al. [22]. It considers ambiguities describing how the system materializes (i.e. displays) and interprets a Visual Sentence, defined as the composition of simple visual elements (graphemes) [22]; these visual elements can be grouped to form structured visual sentences according to the user's actions performed to formulate them. Considering graphemes, which are the lowest level constituent of VLs, it might be ambiguous just as a word may produce a lexical ambiguity as in NL. Structural ambiguities have been classified in: attachment, occurring for example when a label in diagram can be attached to different part of a visual diagram; gap occurring for example when an element is omitted; and analytic occurring when the role of a constituent may be ambiguous. Examples of those classes of visual ambiguities are provided in [23].

The analysis of the literature has underlined that each class of ambiguities implies dealing with the specific features of the input sentence. In this perspective, next sections of this paper address the problem of multimodal ambiguities by a linguistic point of view, defining their classification. Ambiguities are described on the basis of Multimodal Language and Multimodal Sentence notions, defined in section 3.3. Section 4 extends the classifications of NL and VLs ambiguities taking into account the features that appear when combining modal inputs into a multimodal one. The idea is to consider a Multimodal Sentence as a whole according to a holistic point of view.

## 3. Multimodality: Main Concepts

Starting from the hypothesis of adopting a linguistic approach, this section defines the main concepts used to classify and represent multimodal ambiguities features and, it gives the definition of Multimodal Attribute Grammar, Multimodal Sentence and Multimodal Language. The discussion about multimodal ambiguities is based on the definition of Multimodal Attribute Grammar combined with Linear Logic [24], which extends the Classical Logic [25] with the notion of resource and the concept of formulas as resource. Linear Logic relaxes the monotonicity constraints of the Classical Logic and it models changes over time. These features of Linear Logic satisfy some needs deeply connected with multimodality and its feature concerning changing over time. The dynamic nature of Linear Logic combined with a Multimodal Attribute Grammar, are used in this paper in order to provide the definition of Multimodal Sentence and Multimodal Language.

The following sub-sections will provide the definitions of: Multimodal Attribute Grammar (see section 3.1); Production rules of the grammar (see section 3.2); Multimodal Sentence and Multimodal Language (see section 3.3).

### 3.1. Multimodal Attribute Grammar

A Multimodal Attribute Grammar is a context-free and an advanced attribute-based grammar [13], which allows computing derived attributes of non-terminal symbols using computation embedded into the grammar productions. It is defined as follows:

**Def.1** *A Multimodal Attribute Grammar is a triple (G,A,R) where:*
- *G is a context-free grammar defined as a quadruple (T,N,P,S), where T is the set of terminal symbols, N the set of non-terminal symbols, P the set of production rules and S N the start symbol;*
- *A is the collection of attributes of terminal and non-terminal symbols;*
- *R is the collection of semantic rules.*

## 3.2. Production rules of the Multimodal Attribute Grammar

The production rules of the Multimodal Attribute Grammar can be divided in: rules that refer to the construction of the syntax of the grammar $P^g$; rules about the context $P^c$; and temporal rules $P^t$. So the set P of production rules is: P = {$P^g$, $P^c$, $P^t$}. To define the production rules of the proposed method, that will treat both syntax and semantics of the Multimodal Language, the Linear Logic has been chosen. An exhaustive description of Linear Logic is out of scope of this paper, and the Multiplicative Intuitionist fragment of Linear Logic, which uses the multiplicative connective "$\otimes$" (conjunction of hypotheses) and the linear implication "$\cdots \circ$", is sufficient for our purpose. In particular, this work involves the Multimodal Sentence and its interpretation in NL. Therefore, according to its scope, this work uses the production rules of NL, described in [13], for defining the production rules of the syntax of the grammar $P^g$. Some examples of these rules are presented in Table 1 using Linear Logic notations.

**Table 1.** Examples of production rules for Natural Language

| Rules using linear logic notation | Description of the rules |
|---|---|
| s $\cdots \circ$ np $\otimes$ vp | *Sentence* is composed of *noun phrase* and *verb phrase* |
| np $\cdots \circ$ dt $\otimes$ nn $\otimes$ pp | *Noun phrase* is composed of *determiner* and *noun* and *prepositional phrase* |
| np $\cdots \circ$ np $\otimes$ pp | *Noun phrase* is composed of *noun phrase* and *prepositional phrase* |
| np $\cdots \circ$ nn $\otimes$ pp | *Noun phrase* is composed of *noun* and *prepositional phrase* |
| np $\cdots \circ$ det $\otimes$ nn | *Noun phrase* is composed of *determiner* and *noun* |
| np $\cdots \circ$ jj $\otimes$ nn | *Noun phrase* is composed of *adjective* and *noun* |
| np $\cdots \circ$ nn | *Noun phrase* is composed of *noun* |
| vp $\cdots \circ$ vbz $\otimes$ np | *Verb phrase* is composed of *verb (3rd person singular present)* and *noun phrase* |
| vp $\cdots \circ$ vbz $\otimes$ np $\otimes$ pp | *Verb phrase* is composed of *verb (3rd person singular present)* and *noun phrase* and *prepositional phrase* |
| pp $\cdots \circ$ in $\otimes$ np | *prepositional phrase* is composed of *preposition* and *noun phrase* |

The syntax rules of the grammar ($P^g$), the context rules ($P^c$) and the temporal rules ($P^t$) provide both the syntax structure (syntax-tree) and the semantic of the language.

## 3.3. From Terminal Element to the Multimodal Language Definition

The set of terminal symbols T, which belong to the Multimodal Attribute Grammar (Def. 1), contains the building units of the Multimodal Language. These symbols are the terminal elements of Multimodal Sentences for the Multimodal Language. A terminal element contains information about: the modality used to specify the element; the representation of the element in the specific modality; the temporal interval connected with the element where the first number represents the start time and the second number specifies the end time; the syntactic role that the element plays in the Multimodal Sentence; the semantic definition of the element considering its representation according to the modality. In particular the semantic meaning of the element is given considering a domain ontology that provides the conceptual structure of the context. Therefore, a terminal element $E^i$ is defined in [14] as:

**Def.2** *A terminal element $E^i$ is a 5-tuple ($E^i_{mod}$, $E^i_{repr}$, $E^i_{time}$, $E^i_{role}$, $E^i_{concept}$), with:*
- *$E^i_{mod}$: that defines the modality used to create the element $E^i$*
- *$E^i_{repr}$: that defines the representation of the element $E^i$ in the specific modality,*
- *$E^i_{time}$: that defines the temporal interval connected with the element $E^i$,*
- *$E^i_{role}$: the syntactic role that the element $E^i$ plays in the Multimodal Sentence,*
- *$E^i_{concept}$: that specifies the concept name of $E^i$ referred to the conceptual structure of the context.*

For example, let the concept "river" be given and the multimodal input defined both by the sketch and the speech modalities. The representation corresponding to the sketch modality is (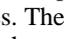), while the representation related to the speech modality is the signal connected with the word river (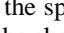 "river"). The element defined by the sketch modality is characterized by the following 5-tuple:
- $E^i_{mod}$= (sketch)
- $E^i_{repr}$= (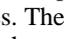 " ")
- $E^i_{time}$= (5, 14)
- $E^i_{role}$= (nn)
- $E^i_{concept}$=(river)

The same element defined by the speech modality, differs from the element defined by the sketch modality in the attribute connected with the modality, in the attribute connected with the representation and in the attribute connected with the temporal interval. In particular, the element defined using the speech modality is characterized by the following 5-tuple:
- $E^i_{mod}$= (speech)
- $E^i_{repr}$= ("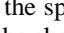 "river" ")
- $E^i_{time}$= (3, 5)
- $E^i_{role}$= (nn)
- $E^i_{concept}$=(river)

When a multimodal composition of terminal elements is ambiguous, it can be associated with more than one syntax-tree, one for each candidate interpretations for the multimodal input expressed with a sentence in NL. All the syntax-trees are combined in a direct acyclic graph, hereafter referred to as syntax-graph. This structure collapses common structures of the different trees associated with the multimodal sentence. In more detail, each terminal node of the syntax-graph and each terminal element of the multimodal attribute grammar matches, and each terminal node includes information about the specific terminal element.

Let W be the Penn Treebank Tag set [26]; a path on the syntax-graph can be defined as:

**Def. 3** *A syntactic (u,v)-path is an ordered sequence of syntactic roles {$w_0$=u, $w_1$, $w_2$,...$w_j$=v} such that $w_i \in W$ for i=0, ...j*

Starting from the introduced concepts, the definition of Multimodal Sentence is here given; it is the grammatical unit containing terminal elements that form functional or perceptual units for the user's input. It is defined as follows:

**Def. 4** *A Multimodal Sentence is a 4-tuple MMS : ( E, syntax-graph, d, int_mms)*
*where :*
- *E is a set of elements $E^i$ for i=1..n, that are terminal elements of the grammar, where n is the number of elements that compose the Multimodal Sentence;*
- *Syntax-graph is a direct acyclic syntactic graph obtained combining all the syntax-trees for modal inputs, and having the terminal elements $E^i$ of the grammar as terminal nodes;*

- *d is the description that defines the meaning of the Multimodal Sentence;*
- *int_mms is the interpretation function that maps the syntax_graph into the description*
- *d: int_mms: (syntax_graph)➔d*

The Multimodal Sentence is used in the Multimodal Language definition.

**Def.5** *A Multimodal Language is a set of Multimodal Sentences*

In order to clarify the given definitions, let the following example of Multimodal Sentence be given: the user says by speech "show this near lake", while she/he is drawing ( ⌒ ) and ( 🔵 ), as described in Figure 2, where the different inputs for the Multimodal Sentence are represented according to their temporal relations. Modal inputs are combined when they satisfy the CloseBy relation - i.e. their associated temporal intervals have at least a common point or they are separated by a number of points less than an established threshold - [27]. The syntax-graph for the Multimodal Sentence of Figure 2 is presented in Figure 3.

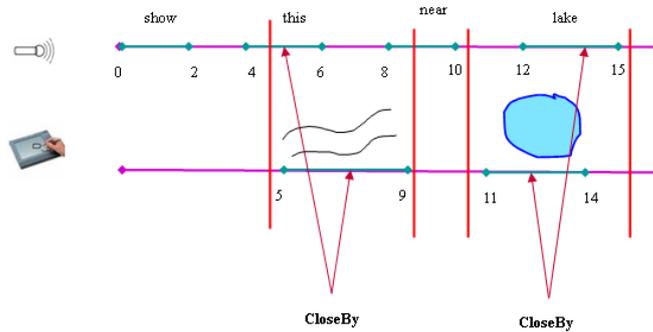

**Figure 2.** Example of Multimodal Sentence with complementary and redundant elements

The syntax-graph (Figure 3), for the Multimodal Sentence of Figure 2, is obtained by parsing the elements of the multimodal sentence by using the rules of Stanford Parser for the Natural Language [28].

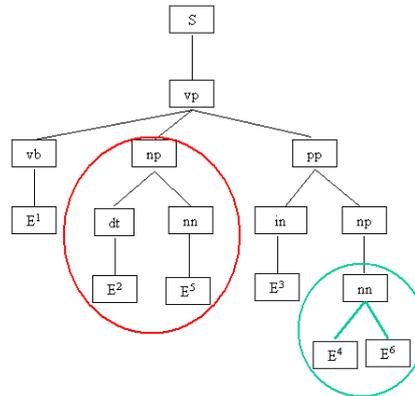

**Figure 3.** Syntax-graph of a Multimodal Sentence with complementary and redundant elements

The elements defined by the speech modality are:

- $E^1$ is! ($E^1_{mod}$=speech) $\otimes$ ! ($E^1_{repr}$=( ⌒ ) "show")) $\otimes$ ! ($E^1_{time}$=(0,2)) $\otimes$ ! ($E^1_{concept}$=(show)) $\otimes$ !($E^1_{role}$=(vb))
- $E^2$ is! ($E^2_{mod}$=speech) $\otimes$ ! ($E^2_{repr}$=( ⌒ ) "this")) $\otimes$ ! ($E^2_{time}$=(4,6)) $\otimes$ ! ($E^2_{concept}$=(deictic)) $\otimes$ !($E^2_{role}$=(dt))
- $E^3$ is! ($E^3_{mod}$=speech) $\otimes$ ! ($E^3_{repr}$ =( ⌒ ) "near")) $\otimes$ ! ($E^3_{time}$=(8,10)) $\otimes$ ! ($E^3_{concept}$=(near)) $\otimes$ !($E^3_{role}$=(in))

- $E^4$ is ! ($E^4_{mod}$=speech) $\otimes$ ! ($E^4_{repr}$ =( "lake")) $\otimes$ ! ($E^4_{time}$=(12,15)) $\otimes$ ! ($E^4_{concept}$=(lake)) $\otimes$ !($E^4_{role}$=(nn))

The elements defined by the sketch modality are:

- $E^5$ is ! ($E^5_{mod}$=sketch) $\otimes$ ! ($E^5_{repr}$ =(")) $\otimes$ ! ($E^5_{time}$=(5,9)) $\otimes$ ! ($E^5_{concept}$=(river)) $\otimes$ !($E^5_{role}$=(nn))

- $E^6$ is ! ($E^6_{mod}$=sketch) $\otimes$ ! ($E^6_{repr}$ =( " ")) $\otimes$ ! ($E^6_{time}$=(11,14)) $\otimes$ ! ($E^6_{concept}$=(lake)) $\otimes$ !($E^6_{role}$=(nn))

The couple of elements E4 and E6 are associated because they are in the CloseBy relation, i.e. their temporal intervals have common points, according to the definition provided in [27]. E4 and E6 express the same concept ("lake"), so they are terminal nodes associated to redundant elements having the same antecedent node (see Figure 3): they have to define redundant elements (i.e. elements referring to the same concept by using two different modalities). Terminal nodes, which have different antecedent nodes, are associated to complementary inputs (E2 and E5 in Figure 3). The two elements of the previous example are complementary because they are in the CloseBy relation and they define two different concepts that must be used to define a specific target concept [27].

The following section uses the given definitions to analyze different multimodal ambiguities and to describe how they are grouped in classes.

## 4. The Multimodal Ambiguities Classification Method

Starting from the ambiguities classification previously described for NL and VLs, this section gives a classification of ambiguities characterizing multimodal interaction processes. This classification has been empirically defined on the basis of a set of 520 multimodal sentences (both ambiguous and un-ambiguous). There are sentences known and classified by the literature containing unimodal ambiguities and sentences with multimodal ambiguities (totally 480), and 40 un-ambiguous sentences. A set of six people, after a tutorial phase, were required to see and hear each multimodal sentence of the set and to annotate if it was ambiguous in their opinion. In this case, it was required to annotate the type of ambiguity (selected between the classes proposed for VLs and NL), if recognized, or two of the possible interpretations. Starting from these interpretations we extracted rules from the common features of the multiple interpretations, producing the empirical classification on the basis of human judgment. The results particularly highlight that multimodal ambiguities arise both, if modal ambiguities are propagated at the multimodal level, and/or if elements, connected with each modality, are correctly interpreted but information referred to different modalities are incoherent at the syntactic (having more than one syntax-tree) or the semantic level (i.e. a multimodal sentence having different meanings). In detail, this section shows the classification of ambiguities we have defined for multimodal interaction.

By observing how people generally communicate using all their five senses both separately and/or in a combined manner, and considering classifications defined for ambiguities in VLs and NL, we have divided multimodal ambiguities into: Semantic ambiguities concerning issues about the meaning of the multimodal input and its components (the elements); and Syntactic ambiguities referring to interpretation problems due to the structure of the multimodal input. As described in the previous sections, the proposed method is based on the Multimodal Attribute Grammar, and the rules about the grammar ($P^g$), the context ($P^c$) and the temporal relations among modalities ($P^t$) are expressed using Linear Logic. Starting from the set P = {$P^g$, $P^c$, $P^t$} of production rules of the Multimodal Language, this work presents their extension by a set of rules $P^a$ that will allow detecting if the Multimodal Sentence is ambiguous and which class of ambiguities appears.

The next sub-sections describe in detail the different classes and present rules we have identified to intercept the different ambiguities in the Multimodal Sentences.

### 4.1. Semantic Ambiguities

This section describes the ambiguities for a multimodal sentence considering its different meanings. In detail, semantic ambiguities are distinguished in lexical, temporal-semantic and target ambiguities.

### 4.1.1. Lexical Ambiguity

A lexical ambiguity deals with the semantics of the elements of the Multimodal Sentence and it appears when the meaning of an element is not clearly identified, as the element has more than one generally accepted meaning. For example, let us suppose the user says using the speech modality "show this in Rome", while she/he simultaneously is drawing the following sketch " ⁓ ". Considering the drawing and the set of elements of the Multimodal Sentence (see Figure 4) and their meanings (according to a defined dictionary), the sketch in Figure 4 can be interpreted both as a river and a street, and the meaning of the user's input is not clearly identified.

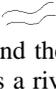

**Figure 4.** Example of Multimodal Sentence

The syntax-graph of the Multimodal Sentence of Figure 4 is presented in Figure 5.

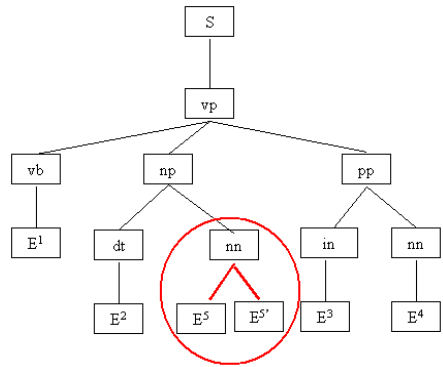

**Figure 5.** Syntax-graph of the user's Multimodal Sentence

The elements defined by speech for the Multimodal Sentence in Figure 4 are:

- $E^1$ is! $(E^1{}_{mod}=speech)$ $\otimes$ ! $(E^1{}_{repr}=(\text{☞ } "show"))$ $\otimes$ ! $(E^1{}_{time}=(0,3))$ $\otimes$ ! $(E^1{}_{concept}=(verb))$ $\otimes !(E^1{}_{role}=(vb))$

- $E^2$ is! $(E^2{}_{mod}=speech)$ $\otimes$ ! $(E^2{}_{repr}=(\text{☞ } "this"))$ $\otimes$ ! $(E^2{}_{time}=(6,8))$ $\otimes$ ! $(E^2{}_{concept}=(deictic))$ $\otimes !(E^2{}_{role}=(dt))$

- $E^3$ is! $(E^3{}_{mod}=speech)$ $\otimes$ ! $(E^3{}_{repr}=(\text{☞ }"in"))$ $\otimes$ ! $(E^3{}_{time}=(12,15))$ $\otimes$ ! $(E^3{}_{concept}=(adverb))$ $\otimes !(E^3{}_{role}=(in))$

- $E^4$ is! $(E^4{}_{mod}=speech)$ $\otimes$ ! $(E^4{}_{repr}=(\text{☞ }"\,Rome"))$ $\otimes$ ! $(E^4{}_{time}=(17,20))$ $\otimes !(E^4{}_{concept}=(city))$ $\otimes !(E^4{}_{role}=(nn))$

While the drawing, defined by the sketch input, can be referred to the two different concepts of river and street, as showed by the $E^5{}_{concep}$ and $E^{5'}{}_{concept}$ below:

- $E^5$ is! $(E^5{}_{mod}=sketch)$ $\otimes$ ! $(E^5{}_{repr}=(\text{🖊 ⁓ }))$ $\otimes$ ! $(E^5{}_{time}=(7,13))$ $\otimes$ ! $(E^5{}_{concept}=(river))$ $\otimes !(E^5{}_{role}=(nn))$

- $E^{5'}$ is! $(E^{5'}{}_{mod}=sketch) \otimes !$ $(E^{5'}{}_{repr} =($ ◠ ◡ $)) \otimes !$ $(E^{5'}{}_{time}=(7,13)) \otimes !$ $(E^{5'}{}_{concept}=(street))$ $\otimes !(E^{5'}{}_{role}=(nn))$

In fact, in this case the alignment of the element $E^2$ with the element $E^5$ detects a lexical ambiguity, because the element $E^5$ can have two different meanings, river ($E^5$) and street ($E^{5'}$), in the context, and the deictic "this" of the element $E^2$ is not useful to disambiguate the meaning. Thereby, a lexical ambiguity is due to the fact that at least one node of the syntax-graph has more than one successor defining different concepts that do not refer to the same semantic concept. The rule that allows identifying this ambiguity is the following:

$E^i$, $E^j$: elements, n: node of the syntax-graph where

$$\exists\ E^i, E^j : elements,\ n : node\ of\ the\ syntax\text{-}graph\ where$$
$$((E^i{}_{concept} \neq E^j{}_{concept})\ \otimes\ (E^i{}_{repr} \equiv E^j{}_{repr})\ \otimes\ (E^i{}_{mod} \equiv E^j{}_{mod})\ \otimes\ (E^i{}_{role} \equiv E^j{}_{role})\ \otimes\ ((E^i,n),\ (E^j,n)\ are$$
$$arcs\ of\ the\ syntax\text{-}graph$$

This rule intercepts lexical ambiguities appearing at the modal level and propagating themselves at the multimodal level.

### 4.1.2. Temporal-Semantic Ambiguity

A temporal-semantic ambiguity appears when two different concepts of two different elements connected with two different input modalities, with a non-empty time intervals intersection or time intervals that are closer in comparison to a threshold value, have the same syntactic role. Therefore, these different elements are terminal nodes having the same parent node in the syntax graph. Let the example of Multimodal Sentence (Figure 6) be given where the user says by speech "this is a river", while she/he draws the following sketch " ◯ ".

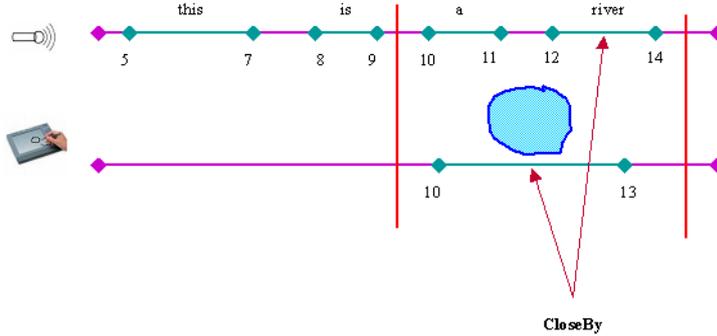

**Figure 6.** Example of input for Multimodal Sentence that defines a temporal-semantic ambiguity

The elements defined by the speech modality are:

- $E^1$ is! $(E^1{}_{mod}=speech) \otimes !$ $(E^1{}_{repr}=($ ◁ ◁ ◁ ⟨ "this")) $\otimes\ !$ $(E^1{}_{time}=(5,7)) \otimes !$ $(E^1{}_{concept}=(deictic))$ $\otimes !(E^1{}_{role}=(dt))$
- $E^2$ is! $(E^3{}_{mod}=speech) \otimes !$ $(E^2{}_{repr} = ($ ⟨ "is")) $\otimes\ !$ $(E^2{}_{time}=(8,9)) \otimes !$ $(E^2{}_{concept}=(is))$ $\otimes !(E^2{}_{role}=(vb))$
- $E^3$ is! $(E^3{}_{mod}=speech) \otimes !$ $(E^3{}_{repr} = ($ ⟨ "a")) $\otimes !$ $(E^3{}_{time}=(10,11)) \otimes !$ $(E^3{}_{concept}=(a))$ $\otimes !(E^3{}_{role}=(dt))$
- $E^4$ is ! $(E^4{}_{mod}=speech) \otimes !$ $(E^4{}_{repr} = ($ ⟨ "river")) $\otimes !$ $(E^4{}_{time}=(12,14)) \otimes !$ $(E^4{}_{concept}=(river))$ $\otimes !(E^4{}_{role}=(nn))$

The element defined by the sketch input is:

- $E^5$ is ! $(E^5{}_{mod}=sketch) \otimes !$ $(E^5{}_{repr} =($ ◠ ◯ $)) \otimes !$ $(E^5{}_{time}=(10,13)) \otimes !$ $(E^5{}_{concept}=(lake))$ $\otimes !(E^5{}_{role}=(nn))$

The syntax-graph of the Multimodal Sentence is presented in Figure 7.

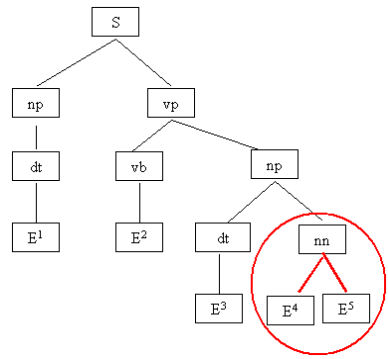

**Figure 7.** Syntax-graph of the Multimodal Sentence that defines a temporal-semantic ambiguity

Also in this case there are two elements that are terminal nodes of the same parent node. They correspond to two different concepts, which are not coherent at the semantic level. Therefore, the rule detecting this class of ambiguity is the following:

$$\exists\ E^i,\ E^j:\ elements,\ n:\ node\ of\ the\ syntax\text{-}graph\ \ where\ ((E^i_{concept} \neq E^j_{concept}) \otimes (E^i_{repr} \neq E^j_{repr}) \otimes$$
$$(E^i_{mod} \neq E^j_{mod}) \otimes\ (E^i_{role} \equiv E^j_{role}) \otimes (E^i_{time}\ CloseBy\ E^j_{time}) \otimes ((E^i,n),\ (E^j,n)\ are\ arcs\ of\ the\ syntax\text{-}\ graph)$$

### 4.1.3. Target Ambiguity

Finally, the target ambiguity appears when the user's focus is not clear. In particular, it appears when at least two elements can be identified as the user's target and they share the same role in the structure of the Multimodal Sentence. Let us suppose, for example, that the user interacts with an interactive map by speech and sketch modalities. By speech, she/he says "show this near school" and, during the same time interval, she/he selects with one only action both the icon of a hotel and the icon of a restaurant by sketch (Figure 8).

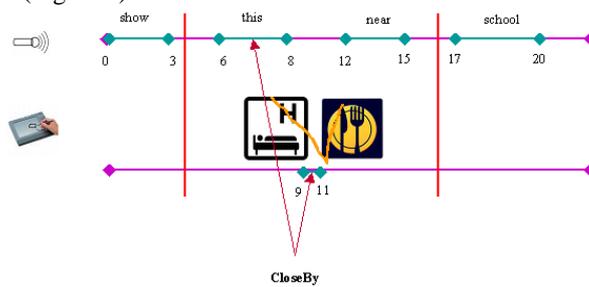

**Figure 8.** Example of input for Multimodal Sentence that defines a target ambiguity

The target ambiguity appears because the user selects two different elements ("hotel" and "restaurant") using the sketch modality (Figure 8) and the selected elements are in CloseBy relation with the deictic "this", provided by speech modality. Therefore the three elements are combined in the interpretation process. This fact implies that the user needs to select only one of the two elements, but the system cannot identify if the user's target is the hotel or the restaurant. This Multimodal Sentence defines the syntax-graph in Figure 9.

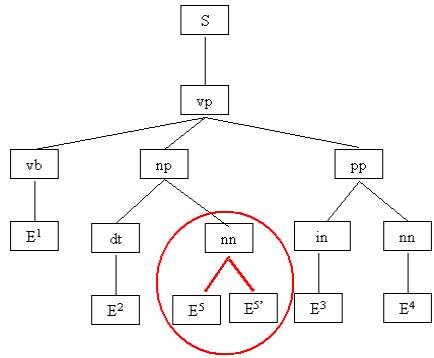

**Figure 9.** Syntax-graph of the Multimodal Sentence that defines a target ambiguity

Elements defined by the speech are:

- $E^1$ is! $(E^1_{mod}=speech) \otimes !$ $(E^1_{repr}=($ "show")) $\otimes !$ $(E^1_{time}=(0,3)) \otimes !$ $(E^1_{concept}=(verb))$ $\otimes !(E^1_{role}=(vb))$

- $E^2$ is! $(E^3_{mod}=speech) \otimes !$ $(E^2_{repr}=($ "this")) $\otimes !(E^2_{time}=(6,8)) \otimes !(E^2_{concept}=(deictic))$ $\otimes !(E^2_{role}=(dt))$

- $E^3$ is! $(E^3_{mod}=speech) \otimes !$ $(E^3_{repr}=($ "near")) $\otimes !$ $(E^3_{time}=(12,15)) \otimes !(E^3_{concept}=(adverb))$ $\otimes !(E^3_{role}=(in))$

- $E^4$ is! $(E^4_{mod}=speech) \otimes !$ $(E^4_{repr}=($ " school")) $\otimes !$ $(E^4_{time}=(17,20)) \otimes !(E^4_{concept}=(city))$ $\otimes !(E^4_{role}=(nn))$

And using the sketch modality the user checks the following elements:

- $E^5$ is ! $(E^5_{mod}=sketch) \otimes !$ $(E^5_{repr}=($)) $\otimes !$ $(E^5_{time}=(9,11)) \otimes !$ $(E^5_{concept}=(hotel))$ $\otimes !(E^5_{role}=(nn))$

- $E^{5'}$ is ! $(E^{5'}_{mod}=sketch) \otimes !$ $(E^{5'}_{repr}=($)) $\otimes !$ $(E^{5'}_{time}=(9,11)) \otimes !$ $(E^{5'}_{concept}=(restaurant)$ $\otimes !(E^{5'}_{role}=(nn))$

The alignment of the element $E^2$ with the elements $E^5$ and $E^{5'}$ detects a target ambiguity due to the fact that, by using sketch modality two different elements, "*hotel*" ($E^5$) and "*restaurant*" ($E^{5'}$), are identified. Similarly to the temporal-semantic ambiguity, a target ambiguity identifies, in the syntax-graph, two terminal nodes associated with two different concepts, which are not coherent at the semantic level (Figure 9) and having the same parent. Unlike the temporal-semantic ambiguity, in this case the identified elements have two different representations and, the rule that allows identifying this ambiguity is the following:

$$\exists \ E^i, E^j: elements, n: node \ of \ the \ syntax\text{-}graph \ where$$
$$((E^i_{concept} \neq E^j_{concept}) \otimes (E^i_{repr} \neq E^j_{repr}) \otimes (E^i_{mod} \equiv E^j_{mod}) \otimes (E^i_{role} \equiv E^j_{role}) \otimes ((E^i,n), (E^j,n) \ are$$
$$arcs \ of \ the \ syntax\text{-}graph))$$

## 4.2. Syntactic Ambiguities

Syntactic ambiguities deal with the structure of a Multimodal Sentence and arise when alternative structures for a given set of elements are generated during the interpretation process. When these ambiguities arise, the role that an element of the language plays is not univocally defined, and the elements of a multimodal sentence (input) can be syntactically combined in more than one way. Considering the syntax-graph, this class of ambiguities appears when a terminal node is not completely defined or when more than one path on the syntax-graph allows reaching the same terminal node. Syntactic ambiguities include gap, analytic and attachment ambiguities that will be detailed in the following subsections.

### 4.2.1. Gap Ambiguity

This section starts analyzing the gap ambiguity that arises when an element of the Multimodal Sentence is omitted. As stated by Futrelle [16], gap ambiguity is frequent in diagrams when labels are omitted. Gap ambiguities can be intercepted using both rules about the grammar of the language and rules about the context. Some examples of gap ambiguity appear when the user specifies an action without specifying the object of the action. Moreover, let us suppose that context rules constrain "to associate each deictic of the Multimodal Sentence to a Multimodal Element". Let us consider a user interacting with a map and saying by speech "Find this near this" and immediately after, considering the time intervals relation showed in Figure 10, she/he draws the sketch " " referring to a lake. The alignment of the elements that compose the user input in the Multimodal Sentence is shown in Figure 10.

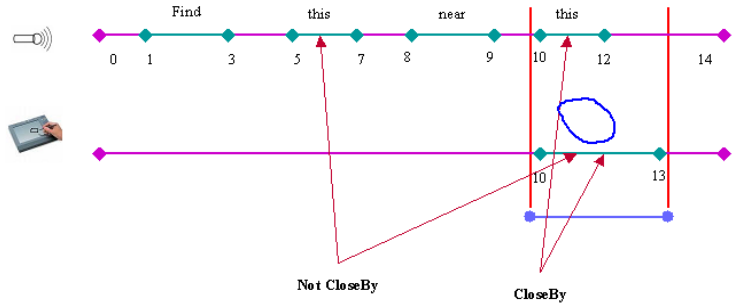

**Figure 10.** Alignment of elements that compose user input by speech and sketch modalities

Figure 11 shows the syntax-graph associated to the Multimodal Sentence of Figure 10. In the example, the element corresponding to the syntactic role (n) (Figure 11) is not defined.

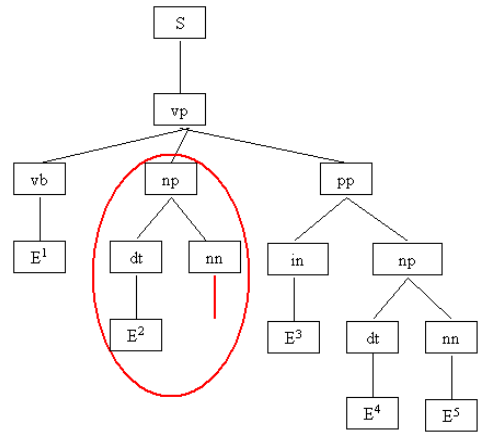

**Figure 11.** Syntax-graph associated to the Multimodal Sentence that defines a gap ambiguity

The Multimodal Sentence is composed by the following elements defined by the speech modalities:

- $E^1$ is! ($E^1_{mod}$=speech) $\otimes$ ! ($E^1_{repr}$= ( "Find" )) $\otimes$ ! ($E^1_{time}$=(1,3)) $\otimes$ ! ($E^1_{concept}$=(verb)) $\otimes$ !($E^1_{role}$=(vb))

- $E^2$ is! ($E^2_{mod}$=speech) $\otimes$ ! ($E^2_{repr}$ = ( "this")) $\otimes$ ! ($E^2_{time}$=(5,7)) $\otimes$ !($E^2_{concept}$=(deictic)) $\otimes$ !($E^2_{role}$=(dt))

- $E^3$ is! ($E^3_{mod}$=speech) $\otimes$ ! ($E^3_{repr}$ = ( "near")) $\otimes$ ! ($E^3_{time}$=(8,9)) $\otimes$ !($E^3_{concept}$=(adverb)) $\otimes$ !($E^3_{role}$=(in))

- $E^4$ is! ($E^4_{mod}$=speech) $\otimes$ ! ($E^4_{repr}$ = ( " this")) $\otimes$ ! ($E^4_{time}$=(10,12)) $\otimes$ !($E^4_{concept}$=(deictic)) $\otimes$ !($E^4_{role}$=( dt))

The element of the Multimodal Sentence for the sketch modality is:

- $E^5$ is ! $(E^5_{mod}=sketch) \otimes ! (E^5_{repr} = (\text{⬭}) \otimes ! (E^5_{time}=(10,13)) \otimes ! (E^5_{concept}=(lake)) \otimes !(E^5_{role}=(nn))$

The syntax graph shows a gap ambiguity because there is an instance of the action ("find"), but there is not an instance of the object of the action ($n$). Moreover, the context rule that imposes to "*associate each deictic of the Multimodal Sentence to a Multimodal Element*" is not satisfied. Considering the previous formalism this ambiguity can be detected using the following rule:

$$\exists \; E^i :: \text{terminal node of the syntax-graph, } n: : \text{vertex s.t. } (E^i \equiv null) \otimes ((E^i,n) \text{ are arcs of the syntax-graph})$$

### 4.2.2. Analytic Ambiguity

A further class of syntactic ambiguities is the analytic one. It arises when the role of an element is not univocally defined in the Multimodal Sentence. In this case the element has more than one possible syntactic role in the Multimodal Sentence.

An example of this ambiguity (widely used in literature for NL) is given by the sentence "The Tibetan history teacher" [15]; it can be interpreted as "(Tibetan history) teacher", i.e. "the teacher of Tibetan history", or "the Tibetan teacher of history". Here, a similar example in a map-based context is presented. Let us suppose that the Multimodal Sentence involves sketch and handwriting modalities and the user says by speech "show Italian river", and immediately after she/he writes the word "name" using the handwriting modality (see Figure 12).

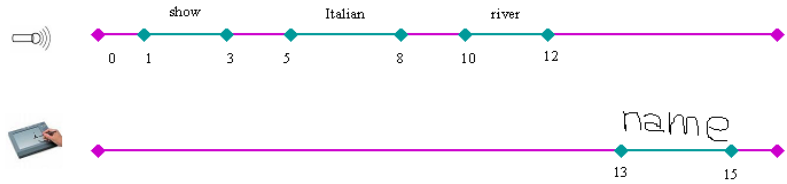

**Figure 12.** User input by sketch and handwriting modalities

This Multimodal Sentence defines the following syntax-graph in Table 2(c).
The elements defined by the speech modality are:

- $E^1$ is! $(E^1_{mod}=speech) \otimes ! (E^1_{repr}=(\text{⌐⟫}$ "show" )) $\otimes ! (E^1_{time}=(1,3)) \otimes ! (E^1_{concept}=(verb)) \otimes !(E^1_{role}=(vb))$
- $E^2$ is! $(E^2_{mod}=speech) \otimes ! (E^2_{repr} = (\text{⌐⟫} \text{ "Italian"})) \otimes ! (E^2_{time}=(5,8)) \otimes !(E^2_{concept}=(adjective)) \otimes !(E^2_{role}=(jj))$
- $E^3$ is! $(E^3_{mod}=speech) \otimes ! (E^3_{repr}=(\text{⌐⟫}"river")) \otimes ! (E^3_{time}=(10,12)) \otimes !(E^3_{concept}=(river)) \otimes !(E^3_{role}=(nn))$
- The element defined by handwriting is:
- $E^4$ is ! $(E^4_{mod}=handwriting) \otimes ! (E^4_{repr}=(\text{✎ name} )) \otimes ! (E^4_{time}=(13,15)) \otimes ! (E^4_{concept}=(name)) \otimes !(E^4_{role}=(nn))$

The syntax-graph obtained by this Multimodal Sentence (Table 2(c)) contains more than one arc reaching the element $E^2$. This element is part of the same *noun phrase* of both the element $E^3$ and $E^4$, because the Multimodal Sentence can be interpreted as: 1) "show the Italian name of the river" (Table 2(a)); and 2) "show the name of the Italian river" (Table 2(b)). The element $E^2$ has two different roles in the syntax-graph (see Table 2(a)-(b)), because there are two different paths that allow reaching this element. Therefore, this ambiguity is produced by the same element $E^j$, which can play two or more different roles in the same Multimodal Sentence.

**Table 2.** Syntax-graph associated to the multimodal sentence that defines an analytic ambiguity

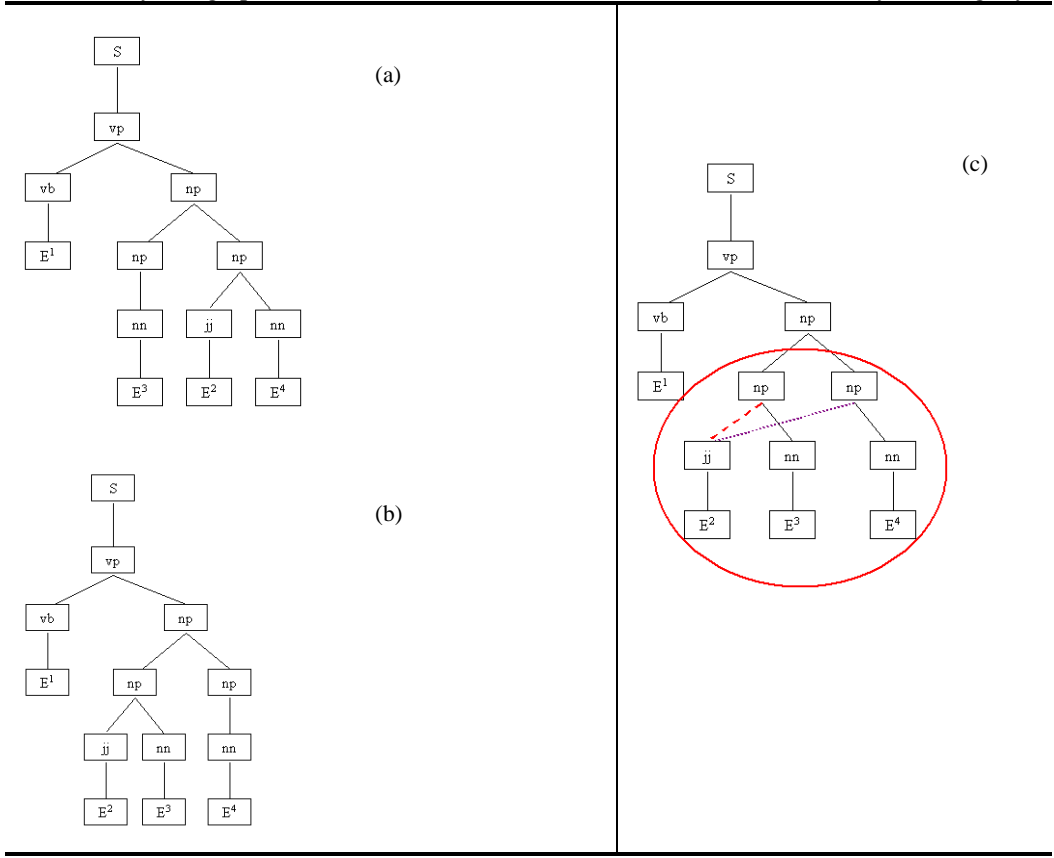

The rule for detecting this class of ambiguity is the following:

$\exists E^j$ : element, n, m : vertexes of the syntax-graph where $(((E^j,n), (E^j,m))$ are paths on the syntax-graph) $\otimes ((E^j,n) \neq (E^j,m)))$

### 4.2.3. Attachment Ambiguity

Considering the attachment ambiguity, it appears when a prepositional phrase (pp) can be legally attached to two different parts of the given sentence. Let us consider now an example of Attachment Ambiguity for a Multimodal Sentence. Let us suppose the user interacts with a map using the sketch and the speech inputs. Using the speech modality the user says "show this near school with garden" and, in the same time interval, she/he draws a house by sketch modality (Figure 13). This Multimodal Sentence defines the syntax-graph in Table 3 (c).

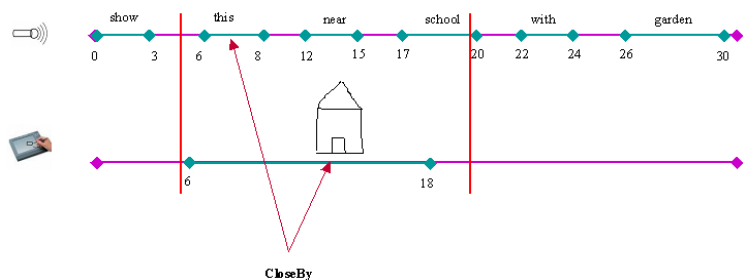

**Figure 13.** Elements defined by the user input

**Table 3.** Syntax-trees and syntax-graph associated to the multimodal sentence that defines an attachment ambiguity

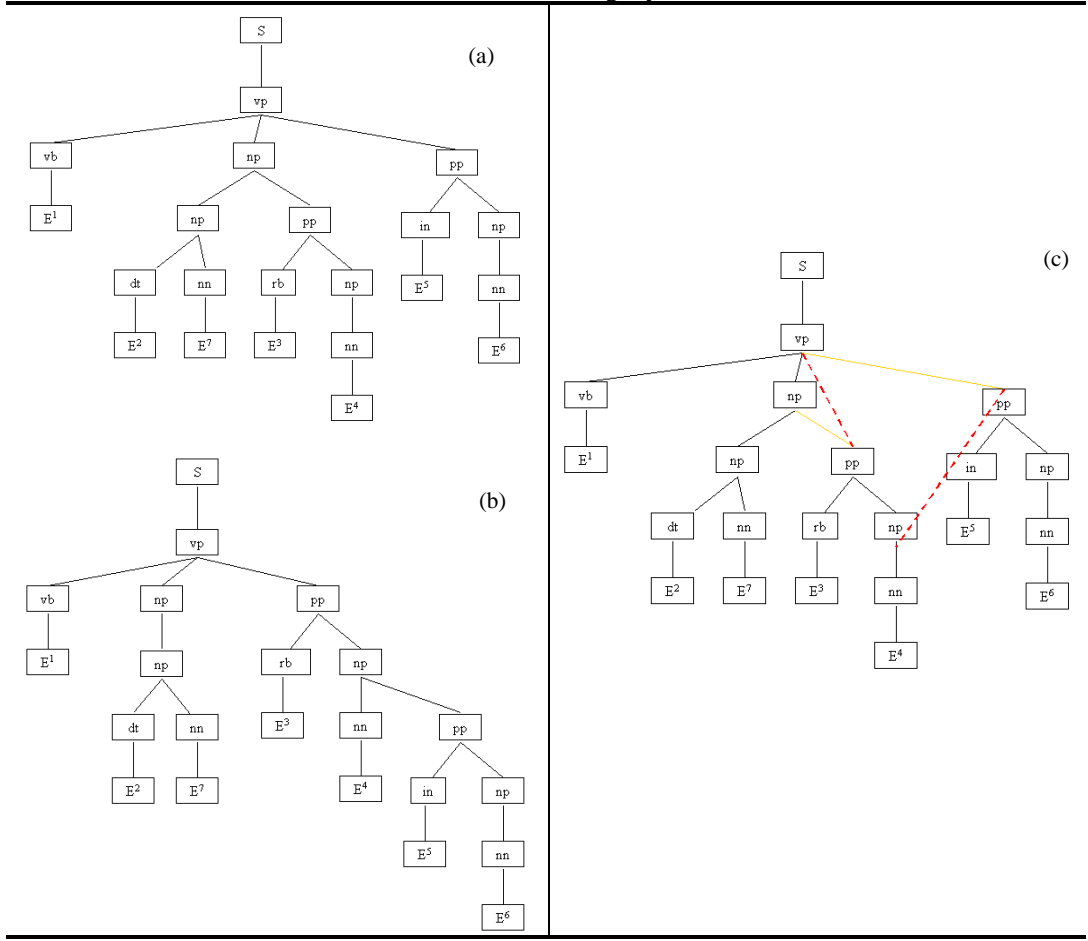

The speech modality elements are:

- $E^1$ is! $(E^1_{mod}=speech) \otimes ! (E^1_{repr} = ($  *"show"*$)) \otimes ! (E^1_{time}=(0,3)) \otimes ! (E^1_{concept}=(verb)) \otimes !(E^1_{role}=(vb))$

- $E^2$ is! $(E^2_{mod}=speech) \otimes ! (E^2_{repr} = ($ *"this"*$)) \otimes ! (E^2_{time}=(6,8)) \otimes ! (E^2_{concept}=(deictic)) \otimes !(E^2_{role}=(dt))$

- $E^3$ is! $(E^3_{mod}=speech) \otimes ! (E^3_{repr} = ($ *"near"*$)) \otimes ! (E^3_{time}=(12,15)) \otimes ! (E^3_{concept}=(adverb)) \otimes !(E^3_{role}=(rb))$

- $E^4$ is! $(E^4_{mod}=speech) \otimes ! (E^4_{repr}= ($ *" school"*$)) \otimes ! (E^4_{time}=(17,20)) \otimes ! (E^4_{concept}=(school)) \otimes !(E^4_{role}=(nn))$

- $E^5$ is! $(E^5_{mod}=speech) \otimes ! (E^5_{repr}= ($ *" with"*$)) \otimes ! (E^5_{time}=(22,24)) \otimes ! (E^5_{concept}=(adverb)) \otimes !(E^4_{role}=( in))$

- $E^6$ is!$(E^6_{mod}=speech) \otimes !(E^6_{repr} = ($ *"garden"*$)) \otimes ! (E^6_{time}=(26,30)) \otimes ! (E^6_{concept}=(garden)) \otimes ! (E^6_{role}=(nn))$

The sketch modality element is:

- $E^7$ is ! $(E^7_{mod}=sketch) \otimes ! (E^7_{repr} =($ $)) \otimes ! (E^7_{time}=(6,18)) \otimes ! (E^7_{concept}=(house)) \otimes !(E^7_{role}=(nn))$

Considering this sentence "*show this (house) near school with garden*", an attachment ambiguity appears because two potential interpretations are defined: 1) in the first syntax-tree "with the garden" is attached to the verb "show", i.e. show this (house) with garden and near school (Table 3(a)); 2) in the second syntax-tree "with the garden" is attached to the element "school", i.e. show this (house) near school with garden (Table 3(b)). Table 3(c) shows that two different paths (----and ——) exist for reaching the prepositional part (pp). Therefore, an attachment ambiguity can be detected using the following rule:

$$\exists\, p, q : \text{syntactic paths, n: node of the syntax-graph where } ((n{=}pp) \otimes (n \in p) \otimes (n \in q) \otimes (p \neq q))$$

### 4.3. Summary of the Multimodal Ambiguities Classification Method

The features of the different classes of multimodal ambiguities are summarised by the decision tree of Figure 14. Others inductive learning algorithms, such as rules-based induction and Bayesian Networks, could have been applied in the classification process, however, we have chosen decision tree-based methods because they allow the recursive partition of data into sub-groups, the selection of an attribute and the formulation of a logical test on attribute.

The decision tree represents a multimodal ambiguities classifier that partitions the set of multimodal ambiguities into six subsets (classes) considering as attributes: the paths on the syntax-graph and the attributes $E^i_{mod}$, $E^i_{repr}$, $E^i_{role}$ and $E^i_{concept}$.

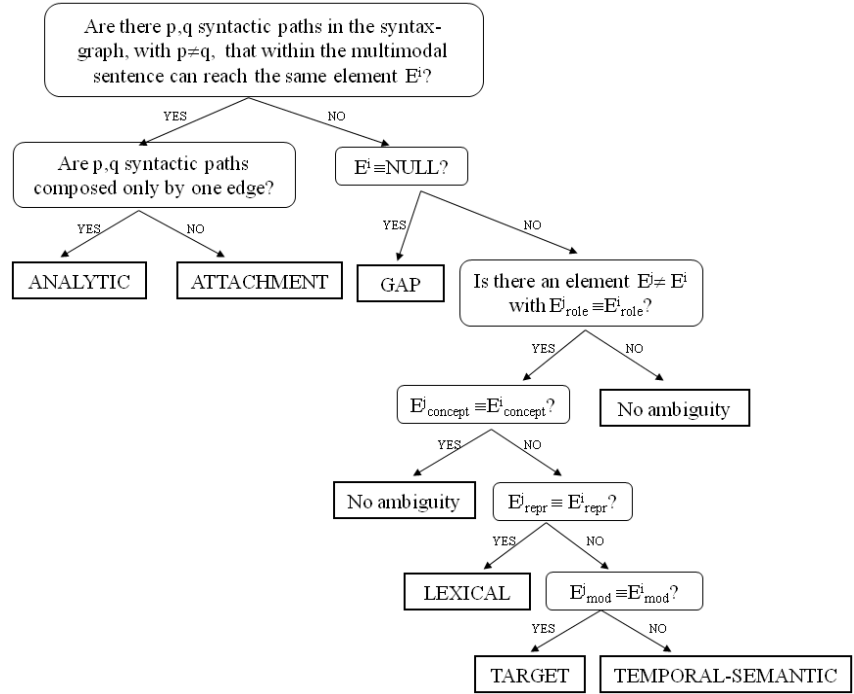

**Figure 14.** Decision tree for the classes of multimodal ambiguities

Table 4 provides an overview of the classes of multimodal ambiguities, a short description of them, how they can be detected on the syntax-graph of the Multimodal Sentence, and rules to intercept them.

**Table 4.** Multimodal ambiguities classes, their descriptions and identification on the syntax-graph and Rules Pᵃ for detecting classes of multimodal ambiguities

| Ambiguity classes | | Description | Identification on the syntax-graph | Rule |
|---|---|---|---|---|
| *Semantic Ambiguities* | Lexical | Arises when one element has more than one generally accepted meaning | There are 2 elements in the syntax-graph that have the same syntactic role, have the same representation as defined in the same modality, but refer to different concepts | $\exists\, E^i, E^j :$ *elements, n : node of the syntax-graph where* $\exists\, ((E^i_{concept} \neq E^j_{concept})\, \otimes (E^i_{repr} \equiv E^j_{repr}) \otimes (E^i_{mod} \equiv E^j_{mod}) \otimes (E^i_{role} \equiv E^j_{role}) \otimes ((E^i,n), (E^j,n)$ *are arcs of the syntax-graph)* |
| | Temporal-Semantic | Arises when two different elements of a multimodal sentence have the same syntactic role but they refer to two different concepts through different modalities | There are 2 elements in the syntax-graph that have the same syntactic role, they have 2 different representations defined by two different modalities, and refer to different concepts | $\exists\, E^i, E^j :$ *elements, n : node of the syntax-graph where* $\exists\, ((E^i_{concept} \neq E^j_{concept}) \otimes (E^i_{repr} \neq E^j_{repr}) \otimes (E^i_{mod} \neq E^j_{mod}) \otimes (E^i_{role} \equiv E^j_{role}) \otimes (E^i_{time}$ *CloseBy* $E^j_{time}) \otimes ((E^i,n), (E^j,n)$ *are arcs of the syntax-graph)* |
| | Target | Arises when the user's focus is not clear | There are 2 elements in the syntax-graph that have the same syntactic role, have different representation but defined in the same modality, but refer to different concepts | $\exists\, E^i, E^j :$ *elements, n : node of the syntax-graph where* $\exists\, ((E^i_{concept} \neq E^j_{concept}) \otimes (E^i_{repr} \neq E^j_{repr}) \otimes (E^i_{mod} \equiv E^j_{mod}) \otimes (E^i_{role} \equiv E^j_{role}) \otimes ((E^i,n), (E^j,n)$ *are arcs of the syntax-graph)* |
| *Syntactic Ambiguities* | Gap | Arises when an element of the multimodal sentence is omitted | There is a terminal node in the syntax-graph that corresponds to a terminal element that has the value NULL | $\exists\, E^i :: $ *terminal node of the syntax-graph, n: : vertex s.t.* $(E^i \equiv null) \otimes ((E^i,n)$ *are arcs of the syntax-graph)* |
| | Analytic | Arises when the syntactic categorization of the element is itself not univocally defined | There are 2 different edges in the syntax-graph that can reach the same element in the syntax-graph | $\exists\, E^j :$ *element, n, m : vertexes of the syntax-graph where* $(((E^j,n), (E^j,m)$ *are paths on the syntax-graph)* $\otimes ((E^j,n) \neq (E^j,m)))$ |
| | Attachment | Arises when a subset of the elements belonging to the sentence can be legally attached to two different parts of the sentence | There are 2 different syntactic paths in the syntax-graph that can reach the same sub-tree in the syntax-graph | $\exists\, p, q :$ *syntactic paths, n: node of the syntax-graph where* $((n=pp) \otimes (n \in p) \otimes (n \in q) \otimes (p \neq q))$ |

## 5. Multimodal Ambiguities Classifier

Identifying the ambiguity class for a Multimodal Sentence is the basis for simplifying the ambiguity solution and optimising the interpretation process. On the basis of the defined classification a software module (the Multimodal Ambiguities Classifier Module of Figure 1) has been implemented. It is an internal module of the overall architecture of the *MultiModal Language Processing framework (M2LP)* [29], which is a platform that aims at managing multimodal communication processes between people and computational systems. The Multimodal Ambiguities Classifier Module identifies the classes of ambiguities connected with the ambiguous multimodal sentences using the rules defined in the section 4.

The following sections provide an example of the use of this module.

### 5.1. Usage of Multimodal Ambiguities Classifier

This section describes an example of how to use the Multimodal Ambiguity Classifier considering the example for lexical ambiguity. The input is provided by an XML file (as shown in Figure 15) representing a Multimodal Sentence transmitted by the *Loader Multimodal Input*.

```xml
<?xml version="1.0"?>
<input>

  <mminput mod="speech" repr="select"  ts="2008-01-15 18:05:20"  te="2008-01-15 18:05:22" conc="select"/>
  <mminput mod="speech" repr="this"  ts="2008-01-15 18:05:24"  te="2008-01-15 18:05:25" conc="this"/>
  <mminput mod="sketch" repr="river_road"  ts="2008-01-15 18:05:23"  te="2008-01-15 18:05:25" conc="river"/>
  <mminput mod="sketch" repr="river_road"  ts="2008-01-15 18:05:23"  te="2008-01-15 18:05:25" conc="road"/>
  <mminput mod="speech" repr="in"  ts="2008-01-15 18:05:27"  te="2008-01-15 18:05:28" conc="in"/>
  <mminput mod="speech" repr="Rome"  ts="2008-01-15 18:05:29"  te="2008-01-15 18:05:30" conc="Rome"/>

  <nlsentence sent="select this river in Rome"/>
  <nlsentence sent="select this road in Rome"/>
</input>
```

**Figure 15.** Example of XML file connected with the ambiguous Multimodal Sentence

For each ambiguous Multimodal sentence the system starts to analyze the multimodal input. The syntax classifier produces the syntax-graph connected with the selected ambiguous multimodal input and it returns the class of ambiguity (in this case the *lexical ambiguity class*). Figure 16 shows the detected class of ambiguity, the syntax graph and the specific problem to solve. In the example shown in Figure 16, the element NN has the two different meanings (road and river) according to the specified context; in fact, there is a node of the syntax-graph that has two successors that refer to the two different concepts road and river.

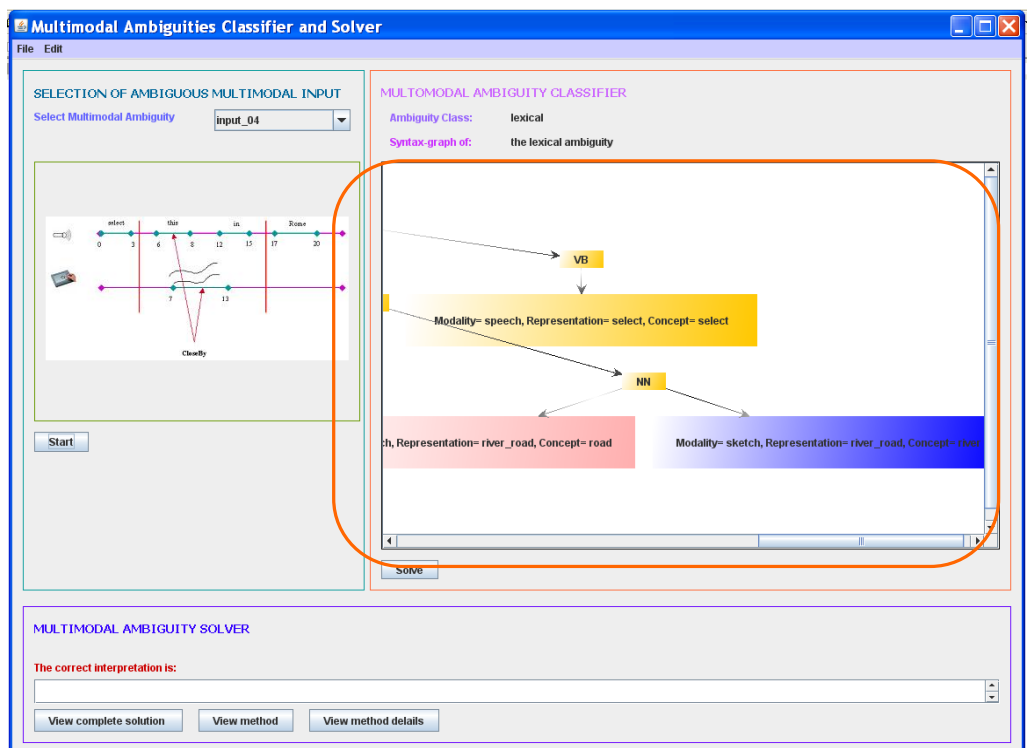

**Figure 16.** Syntax-graph and ambiguity class connected with the ambiguous input

The information about the ambiguous sentence, the classified ambiguity and the associated syntax-graph are transferred to the Multimodal Ambiguities Solution module [14], as Figure 1 shows, in order to provide an unambiguous interpretation of the multimodal sentence.

## 6. Evaluation of Multimodal Ambiguities Classifier Performance

The evaluation process of the Multimodal Ambiguities Classifier consists on analyzing its performances in correctly assigning a multimodal sentence to the ambiguity class.

The test set of 520 Multimodal Sentences, used to evaluate the Multimodal Ambiguities Classifier, is composed of 480 ambiguous Multimodal Sentences and 40 un-ambiguous Multimodal Sentences. Those sentences have been used both for the syntactic and the semantic test set. The set of un-ambiguous sentences has been introduced in order to test the system robustness. The set of 480 ambiguous Multimodal Sentences consists of: 240 multimodal sentences that present a semantic multimodal ambiguity (80 lexical, 80 temporal-semantic and 80 target ambiguities); 240 multimodal sentences that contain a syntactic multimodal ambiguity (80 gap, 80 analytic and 80 attachment ambiguities); and 40 multimodal sentences that are un-ambiguous. The Multimodal Ambiguity Classifier will receive in input the XML files of the Multimodal Sentences defined in the test set, but it doesn't know the class of ambiguity of each sentence of the test set (which represents the expected class of ambiguity). Starting from this consideration, it is possible to calculate the performance in terms of accuracy of the classification process of multimodal ambiguities considering, for each Multimodal Sentence, how the detected class of ambiguity matches the expected class of ambiguity (obtained on the basis of the human judgment expressed in the definition of the classification, as described in Section 4). Table 5 presents some of the ambiguous multimodal inputs used to test the Multimodal Ambiguities Classifier module, their classes of ambiguity and some of their interpretations.

**Table 5.** Examples of inputs for the testing process

| Ambiguity Class | | Examples of multimodal ambiguous sentences of the test set | Possible Interpretations in NL |
|---|---|---|---|
| *Semantic* | *Lexical* | 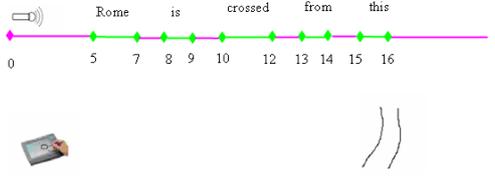 | - Rome is crossed from this river<br><br>- Rome is crossed from this road |
| | *Temporal-Semantic* | 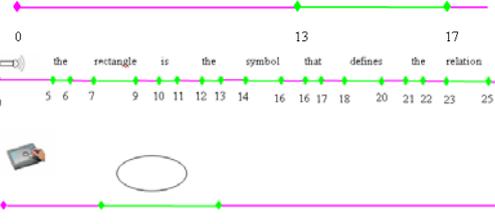 | - The rectangle is the symbol that defines the relation<br><br>- The oval is the symbol that defines the relation |
| | *Target* | 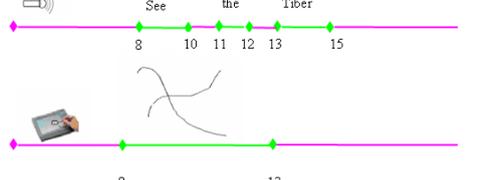 | - See the Tiber river1<br><br>- See the Tiber river2 |
| *Syntactic* | *Gap* | 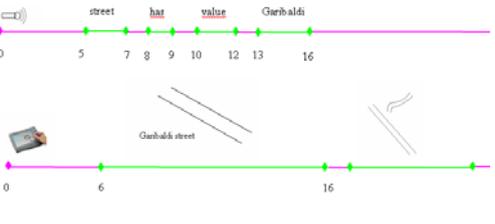 | - Street that has value Garibaldi arrives close to the river (….) |
| | *Analytic* | 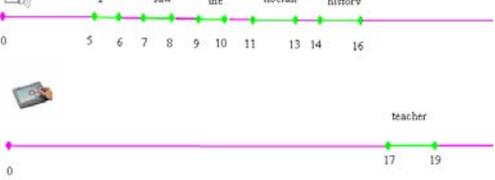 | - I saw the Tibetan teacher of history<br><br>- I saw the teacher of Tibetan history |
| | *Attachment* | 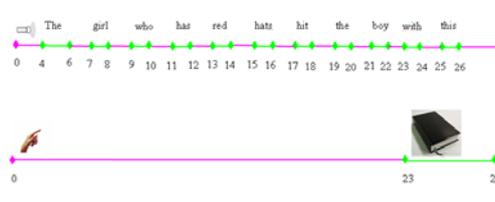 | - The girl who has red hats hit the boy with this book<br><br>- The girl who has red hats hit, the boy with this book |

The classification accuracy of the Multimodal Ambiguities Classifier measures 94.6% for semantic ambiguities, and 92.1% for syntactic ambiguities. In detail, considering the test set, the module has correctly classified 265 examples for semantic multimodal ambiguities, and 258 examples for syntactic multimodal ambiguities (see Figure 17). The un-ambiguous sentences have always been correctly identified.

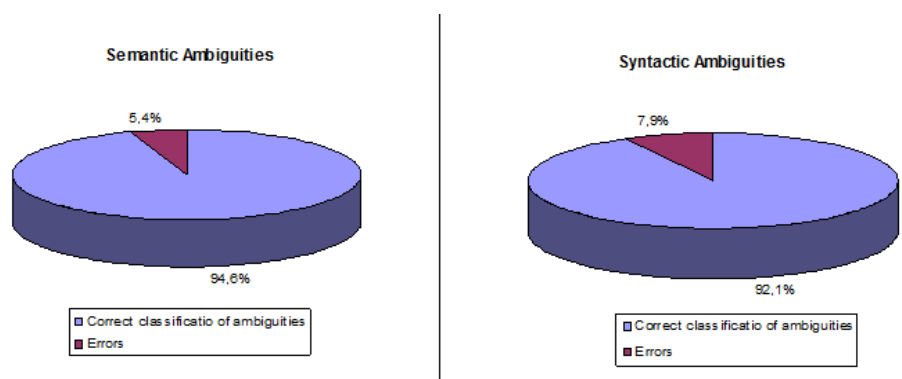

**Figure 17.** Measures of the classification accuracy of the Multimodal Ambiguities Classifier

## 7. Conclusion

This paper has addressed the problem of ambiguity that characterizes multimodal communication processes by a linguistic perspective, proposing their classification. Critical issues characterizing ambiguities and methods provided by the literature have been identified. In particular, an overview of some relevant studies on ambiguities related with the NL and VLs has been carried out and, the most relevant results of these studies have been empirically extended for Multimodal Languages. A set of rules has been defined in order to typify the classes of multimodal ambiguities, coherently with the notions of Multimodal Attribute Grammar, terminal element of the Multimodal Attribute Grammar, Multimodal Sentence and Multimodal Language. The proposed set of rules has been defined on the basis of: 1) the existing studies on VLs and NL, and the observation of some ambiguities in multimodal languages, and 2) their extension considering a set of associations between each ambiguity class and the detected multimodal ambiguities. The defined classification has divided multimodal ambiguities into Semantic and Syntactic ambiguities. The Semantic ambiguities have been classified into: lexical ambiguity; temporal-semantic ambiguity; and target ambiguity. Syntactic ambiguities have been divided into: gap ambiguity; analytic ambiguity; and attachment ambiguity. The results obtained for classifying multimodal ambiguities have been implemented in the design and development of the Multimodal Ambiguities Classifier module. This module has been validated and its evaluation has provided a good level of accuracy of multimodal ambiguities classification. In fact, it has reached a classification accuracy of 94.6% for the semantic ambiguities, and 92.1% for the syntactic ambiguities. The experimental results encourage us to tackle further improvements and tests to the proposed method. For example, a more advanced prototype will be provided using a wide corpus of Multimodal Sentences.

Future works will investigate on how other inductive learning algorithms, and particularly genetic algorithms, are able to classify but also to deal with a possible evolution of such classes.